\documentclass[conference]{IEEEtran}

\ifCLASSOPTIONcompsoc

  \usepackage[nocompress]{cite}
\else
  \usepackage{cite}
\fi
\ifCLASSINFOpdf
  
\else
  
\fi

\usepackage{float}
\usepackage{soul, color, acronym, multirow, amsmath, balance, amssymb}
\usepackage[pdftex]{graphicx}
\usepackage[justification=centering]{caption}
\usepackage{tikz, rotating, float, epsfig, epstopdf, subcaption,enumitem}
\usepackage{booktabs,textcomp,url}
\usepackage{pbox}
\usepackage{listings}
\usepackage{authblk}
\usepackage{pifont}
\usepackage{array}
\usepackage{soul}
\newcolumntype{C}[1]{>{\centering\arraybackslash}m{#1}}
\usepackage{amsmath}
\usepackage{optidef}

\acrodef{PER}{Packet Erasure Rate}
\acrodef{PLR}{Packet Loss Rate}
\acrodef{PRR}{Packet Received Ratio}
\acrodef{BER}{Bit Error Rate}
\acrodef{RTT}{Round-Trip Time}
\acrodef{TCP}{Transmission Control Protocol}
\acrodef{UDP}{User Datagram Protocol}
\acrodef{AIMD}{Additive Increase Multiplicative Decrease}
\acrodef{PEP}{Performance Enhancing Proxy}
\acrodef{cwnd}{congestion window}
\acrodef{STP}{Satellite Transport Protocol}
\acrodef{BDP}{Bandwidth-Delay Product}
\acrodef{QoS}{Quality of Service}
\acrodef{LoS}{Line of Sight}
\acrodef{NLoS}{Non Line of Sight}
\acrodef{BLoS}{Beyond Line of Sight}
\acrodef{QoE}{Quality of Experience}
\acrodef{ACM}{Adaptive Coding and Modulation}
\acrodef{DAMA}{Demand Assignment Multiple Access}
\acrodef{VANET}{Vehicular Ad-Hoc Network}
\acrodef{RA}{Random Access}
\acrodef{CRA}{Contention Resolution ALOHA}
\acrodef{SA}{Slotted ALOHA}
\acrodef{DSA}{Diversity Slotted ALOHA}
\acrodef{OBU}{On-Board Unit}
\acrodef{CRDSA}{Contention Resolution Diversity Slotted ALOHA}
\acrodef{SIC}{Successive Interference Cancellation}
\acrodef{ARQ}{Automatic Repeat reQuest}
\acrodef{SC-ARQ}{Selective-Coded ARQ}
\acrodef{SR-ARQ}{Selective-Repeat ARQ}
\acrodef{IRSA}{Irregular Repetition Slotted ALOHA}
\acrodef{CGC}{Complementary Ground Component}
\acrodef{GCS}{Ground Control Station}
\acrodef{RSU}{Road Side Unit}
\acrodef{ACK}{Acknowledgment}
\acrodef{NACK}{Negative Acknowledgment}
\acrodef{DVB-SH}{Digital Video Broadcasting - Satellite Services to Handhelds}
\acrodef{DVB-H}{Digital Video Broadcasting - Handheld}
\acrodef{DVB-RCS2}{Digital Video Broadcasting - Return Channel via Satellite, II generation}
\acrodef{SACK}{Selective Acknowledgment}
\acrodef{SNACK}{Selective Negative Acknowledgment}\acrodef{SNACK}{Selective Negative Acknowledgment}
\acrodef{MOS}{Mean Opinion Score}
\acrodef{SNIR}{Signal to Noise plus Interference Ratio}
\acrodef{SCPS-TP}{Space Communications Protocol Specifications - Transport Protocol}
\acrodef{CCSDS}{Consultative Committee for Space Data Systems}
\acrodef{ESA}{European Space Agency}
\acrodef{NASA}{National Aeronautics and Space Administration}
\acrodef{BSM}{Broadband Satellite Multimedia}
\acrodef{RLNC}{Random Linear Network Coding}
\acrodef{NC}{Network Coding}
\acrodef{FIFO}{First In, First Out}
\acrodef{FCFS}{First Come, First Served}
\acrodef{BLER}{Block Error Rate}
\acrodef{GEO}{geosynchronous}
\acrodef{LEO}{Low Earth Orbit}
\acrodef{FTP}{File Transfer Protocol}
\acrodef{CRC}{Cyclic Redundancy Check}
\acrodef{MAC}{Media Access Control}
\acrodef{PHY}{Physical layer}
\acrodef{HTTP}{Hypertext Transfer Protocol}
\acrodef{ISP}{Internet Service Provider}
\acrodef{MSS}{Maximum Segment Size}
\acrodef{BIC}{Binary Increase Congestion control}
\acrodef{AQM}{Active Queue Management}
\acrodef{XCP}{eXplicit Control Protocol}
\acrodef{ECN}{Explicit Congestion Notification}
\acrodef{CA}{Congestion Avoidance}
\acrodef{RED}{Random Early Detection}
\acrodef{TD}{Triple-Duplicate}
\acrodef{TO}{TimeOut}
\acrodef{IP}{Internet Protocol}
\acrodef{WMN}{Wireless Mesh Network}
\acrodef{ssthresh}{Slow-Start threshold}
\acrodef{MPE-IFEC}{Multi Protocol Encapsulation - Inter-burst Forward Error Correction}
\acrodef{FEC}{Forward Error Correction}
\acrodef{ML}{Maximum Likelihood}
\acrodef{CRC}{Cyclic Redundancy Check}
\acrodef{P2P}{Peer-to-Peer}
\acrodef{FMT}{Fade Mitigation Technique}
\acrodef{SGD}{Smart Gateway Diversity}
\acrodef{NCC}{Network Control Centre}
\acrodef{ModCod}{Modulation and Coding}
\acrodef{FIFO}{First-In-First-Out}
\acrodef{WRR}{Weighted Round Robin}
\acrodef{WFQ}{Weighted Fair Queuing}
\acrodef{NS}{Network Simulator}
\acrodef{GSE}{Generic Stream Encapsulation}
\acrodef{PDF}{Probability Density Function}
\acrodef{CDF}{Cumulative Density Function}
\acrodef{CoV}{Coefficient of Variation}
\acrodef{MSC}{Message Sequence Chart}
\acrodef{ESA}{European Space Agency}
\acrodef{LIU}{Lebanese International University}
\acrodef{TUM}{Technical University of Munich}
\acrodef{MSCE-CS}{Master of Science in Communications Engineering - Communications Systems}
\acrodef{DLR}{German Aerospace Center}
\acrodef{NOS}{Network Operating System}
\acrodef{NFs}{Network Functionalities}
\acrodef{NC-SGD}{Network Coding for SGD}
\acrodef{ATSP}{Advanced Transport Satellite Protocol}
\acrodef{STP}{Satellite Transport Protocol}
\acrodef{WMN}{Wireless Mesh Network}
\acrodef{SNR}{Signal-to-Noise Ratio}
\acrodef{SINR}{Signal-to-Interference-plus-Noise Ratio}
\acrodef{LMS}{Land Mobile Satellite}
\acrodef{LTE}{Long-Term Evolution}
\acrodef{M2M}{machine-to-machine}
\acrodef{IoT}{Internet of Things}
\acrodef{IoRT}{Internet of Remote Things}
\acrodef{IoST}{Internet of Space Things}
\acrodef{MIoT}{Multimedia Internet of Things}
\acrodef{RA}{Random Access}
\acrodef{UAV}{Unmanned Aerial Vehicle}
\acrodef{UTV}{Unmanned Terrestrial Vehicle}
\acrodef{UAS}{Unmanned Aerial System}
\acrodef{FANET}{Flying Ad-Hoc Network}
\acrodef{MANET}{Mobile Ad-Hoc Network}
\acrodef{VANET}{Vehicle Ad-Hoc Network}
\acrodef{C2}{Command and Control}
\acrodef{DTN}{Delay Tolerant Network}
\acrodef{COTS}{Commercial Off-the-Shelf}
\acrodef{IETF}{Internet Engineering Task Force}
\acrodef{CoAP}{Constrained Application Protocol}
\acrodef{MQTT}{Message Queuing Telemetry Transport}
\acrodef{URI}{Uniform Resource Identifier}
\acrodef{PUB/SUB}{Publish / Subscribe}
\acrodef{RCST}{Return Channel Satellite Terminal}
\acrodef{TDMA}{Time Division Multiple Access}
\acrodef{FDMA}{Frequency Division Multiple Access}
\acrodef{TCDMA}{Turbo Code Division Multiple Access}
\acrodef{PDMA}{Power Division Multiple Access}
\acrodef{WSN}{Wireless Sensor Network}
\acrodef{REST}{Representational State Transfer}
\acrodef{EDGE}{Enhanced Data rates for GSM Evolution}
\acrodef{UMTS}{Universal Mobile Telecommunications System}
\acrodef{LTE}{Long-Term Evolution}
\acrodef{E2E}{End-to-End}
\acrodef{3WHS}{Three-way Handshake}
\acrodef{SCADA}{Supervisory Control And Data Acquisition}
\acrodef{SOA}{Service-Oriented Architecture}
\acrodef{WebRTC}{Web Real-Time Communications}
\acrodef{fps}{frames per second}
\acrodef{SSIM}{Structural SIMilarity}
\acrodef{PSNR}{Peak Signal-to-Noise Ratio}
\acrodef{RPi}{Raspberry Pi}
\acrodef{NFV}{Network Function Virtualization}
\acrodef{OPEX}{Operating Expenditures}
\acrodef{CAPEX}{Capital Expenditures}
\acrodef{MIMO}{multiple-input and multiple-output}
\acrodef{SDN}{Software Defined Networking}
\acrodef{MTC}{Machine-type Communications}
\acrodef{mMTC}{Massive Machine-type Communications}
\acrodef{HTC}{Human-type Communications}
\acrodef{D2D}{device to device}
\acrodef{IIoT}{industrial IoT}
\acrodef{LPWAN}{Low-Power Wide-Area Network}
\acrodef{CPS}{Cyber-Physical System}
\acrodef{ICT}{Information and Communication Technologies}
\acrodef{SDR}{Software Defined Radio}
\acrodef{GPS}{Global Positioning System}
\acrodef{H2M}{human-to-machine}
\acrodef{MC}{Machine}
\acrodef{LA}{Local Area}
\acrodef{WA}{Wide Area}
\acrodef{HAP}{High Altitude Platform}
\acrodef{LAP}{Low Altitude Platform}
\acrodef{RAN}{Radio Access Network}
\acrodef{EIRP}{Equivalent Isotropically Radiated Power}
\acrodef{LT}{Logistics and Transportations}
\acrodef{BVLoS}{Beyond Visual Line of Sight}
\acrodef{BLoS}{Beyond Line of Sight}
\acrodef{VLoS}{Visual Line of Sight}
\acrodef{UTM}{Unmanned Aircraft System Traffic Management}
\acrodef{eNB}{Evolved Node B}
\acrodef{CDN}{Content Distribution Network}
\acrodef{ITU}{International Telecommunication Union}
\acrodef{SIN}{Space Information Network}
\acrodef{DTA}{Delay Tolerant Application}
\acrodef{DSA}{Delay Sensitive Application}
\acrodef{DTN}{Delay Tolerant Networking}
\acrodef{ISL}{Inter-Satellite Link}
\acrodef{TFRC}{TCP Friendly Rate Control}
\acrodef{RACH}{Random Access CHannel}
\acrodef{LPWAN}{Low-Power Wide Area Network}
\acrodef{VRT}{Variable Rate Technology}
\acrodef{HAP}{High Altitude Platform}
\acrodef{LALE}{Low Altitude Long Endurance}
\acrodef{LASE}{Low Altitude Short Endurance}
\acrodef{MALE}{Medium Altitude Long Endurance}
\acrodef{HALE}{High Altitude Long Endurance}
\acrodef{OTA}{Over The Air}
\acrodef{TLS}{Transport Layer Security}
\acrodef{SSL}{Secure Sockets Layer}
\acrodef{MEC}{Mobile Edge Computing}
\acrodef{GIS}{Geographic Information System}
\acrodef{GSM}{Global System for Mobile}
\acrodef{GPRS}{General Packet Radio Service}
\acrodef{PF}{Precision Farming}
\acrodef{SF}{Smart Farming}
\acrodef{SAR}{Syntethic Aperture Radar}
\acrodef{EO}{Earth Observation}
\acrodef{GNSS}{Global Navigation Satellite System}
\acrodef{EGNOS}{European Geostationary Navigation Overlay Service}
\acrodef{API}{Application Programming Interface}
\acrodef{RTP}{Real-time Transport Protocol}
\acrodef{RTCP}{Real-time Transport Control Protocol}
\acrodef{RR}{Receiver Report}
\acrodef{RB}{Round Robin}
\acrodef{WRR}{Weighted Round Robin}
\acrodef{ULFEC}{Upper Layer Forward Error Correction}
\acrodef{IC}{Interference Cancellation}
\acrodef{NIC}{Network Interface Card}
\acrodef{SCTP}{Stream Control Transmission Protocol}
\acrodef{ULPFEC}{Uneven Level Protection FEC}
\acrodef{DTMC}{Discrete Time Markov Chain}
\acrodef{GoP}{Group of Pictures}
\acrodef{NAT}{Network Address Translation}
\acrodef{NTN}{Non-Terrestrial Network}
\acrodef{MDP}{Markov Decision Process}
\acrodef{URLLC}{Ultra-Reliable and Low Latency Communications}
\acrodef{eMBB}{enhanced Mobile BroadBand}
\acrodef{VR}{Virtual Reality}
\acrodef{HTS}{High Throughput Satellite}
\acrodef{SN}{Sequence Number}
\acrodef{PAN}{Path-Aware Networking}
\acrodef{E2E}{End-to-End}
\acrodef{CSI}{Channel State Information}
\acrodef{PoC}{Proof of Concept}
\acrodef{AC}{Actor-Critic}
\acrodef{RL}{Reinforcement Learning}
\acrodef{UE}{User Equipment}
\acrodef{DQ}{Deep-Q}
\acrodef{NN}{Neural Network}
\acrodef{HIL}{Human In the Loop}

\acrodefplural{eNB}[eNBs]{Evolved Nodes B}
\acrodefplural{NIC}[NICs]{Network Interface Cards}
\acrodefplural{RSU}[RSUs]{Road Side Units}
\acrodefplural{VANET}[VANETs]{Vehicular Ad-Hoc Networks}
\acrodefplural{OBU}[OBUs]{On-Board Units}
\acrodefplural{UAV}[UAVs]{Unmanned Aerial Vehicles}
\acrodefplural{FANET}[FANETs]{Flying Ad-Hoc Networks}
\acrodefplural{MANET}[MANETs]{Mobile Ad-Hoc Networks}
\acrodefplural{VANET}[VANETs]{Vehicle Ad-Hoc Networks}
\acrodefplural{UAS}[UASs]{Unmanned Aerial Systems}
\acrodefplural{HAP}[HAPs]{High Altitude Platforms}
\acrodefplural{LALE}[LALEs]{Low Altitude Long Endurance}
\acrodefplural{LASE}[LASEs]{Low Altitude Short Endurance}
\acrodefplural{MALE}[MALEs]{Medium Altitude Long Endurance}
\acrodefplural{HALE}[HALEs]{High Altitude Long Endurance}
\acrodefplural{RCST}[RCSTs]{Return Channel Satellite Terminals}

\begin{document}


%

\title{Actor-Critic Scheduling for Path-Aware Air-to-Ground Multipath Multimedia Delivery}

\author[*1]{Achilles Machumilane}
\author[1,2]{Alberto Gotta}
\author[1,2]{Pietro Cassar\'a}
\author[1]{Claudio Gennaro}
\author[1]{Giuseppe Amato}
\affil[*]{Department of Information Engineering, University of Pisa-achilles.machumilane@phd.unipi.it}
\affil[1]{Institute of Information Science and Technologies (ISTI), CNR, Pisa - e-mails: \{name.surname\}@isti.cnr.it}
\affil[2]{CNIT - National Inter-University Consortium for Telecommunications}

\maketitle
\begin{abstract}
\ac{RL} has recently found wide applications in network traffic management and control because some of its variants do not require prior knowledge of network models. In this paper, we present a novel scheduler for real-time multimedia delivery in multipath systems based on an \ac{AC} \ac{RL} algorithm. 
We focus on a challenging scenario of real-time video streaming from an \ac{UAV} using multiple wireless paths. The scheduler acting as an \ac{RL} agent learns in real-time the optimal policy for path selection, path rate allocation and  redundancy estimation for flow protection. The scheduler, implemented as a module of the GStreamer framework, can be used in real or simulated settings.  The simulation results show that our scheduler can target a very low loss rate at the receiver by dynamically adapting in real-time the scheduling policy to the path conditions without performing training or relying on prior knowledge of network channel models. 
\end{abstract}


\section{Introduction}
\label{sec:introduction}
Services that make use of real-time streaming available to
of a communications infrastructure are becoming increasingly		mobile users of a communication infrastructure for scenarios
popular in scenarios such as vehicular networks, flying ad-		such as vehicular networks, flying ad-hoc networks or under-
hoc networks, or underwater networks. In these scenarios,		water networks are increasingly popular. In these scenarios, multipath techniques are emerging as key strategies to improve channel reliability and availability for multimedia transmission, especially in wireless networks. Multipath transmission is a technique leveraging network diversity in which the same traffic is shared or replicated over multiple paths. In this way, data lost on one path has high probability of being recovered from other paths.  The receiver reassembles and reorders the data from different paths and forwards the data to the application. Multipath transmission has several advantages: \textit{(i)} it can improve link availability and reliability, \textit{(ii)} it can optimise transmission by maximizing throughput or minimizing delay, 
and \textit{(iii)} it allows for path aggregation and load balancing. 
Consequently, it has the potential to improve both \ac{QoS} and user-perceived \ac{QoE} in real-time multimedia applications. Among the scenarios stated above, those involving \acp{UAV} can largely benefit from multipath transmissions, mainly when operated over wireless networks, without \ac{LoS} \cite{bacco2017uavs, bacco2017survey}. 

A critical aspect of  multipath transmission is packet scheduling. A multipath scheduler must perform three main tasks: \textit{(i)} path selection, \textit{(ii)} packet selection and \textit{(iii)} packet/flow protection \cite{afzal2019holistic}.

A multipath scheduler can address these tasks by using the \ac{CSI} to perform optimal resource allocation on each path. Performing these tasks is more challenging in wireless networks because of the heterogeneity and time-variability of the paths, in terms of available bandwidth, delay, cost, and congestion level.
This work aims to address the optimal allocation of the data streaming over multiple paths, developing a Path-Aware Dynamic Multipath schEduler (PADME) based on an \ac{AC} algorithm, called AC-PADME, where the core of the scheduler is an interleaved \ac{WRR} mechanism to be implemented on the on-board unit of a \ac{UAV}. 

\begin{figure}
\centering
    \includegraphics[width=1\columnwidth]{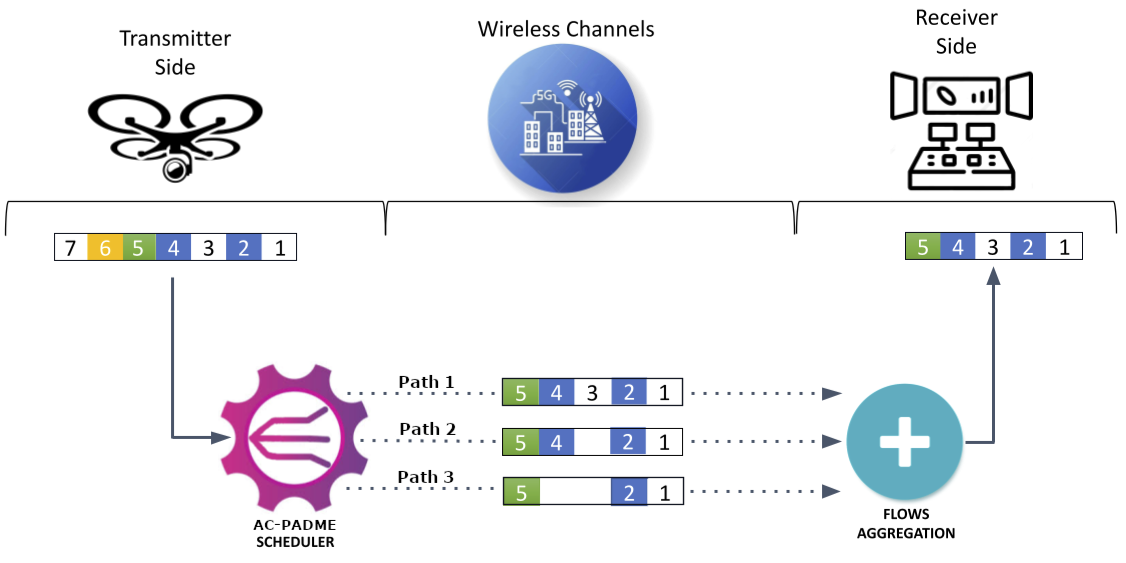}
    \caption{A multimedia stream is sent from a \ac{UAV} to a \ac{GCS} via multiple paths through AC-PADME scheduler that selects the paths to be used and the necessary redundancy rate.}
    \label{fig:scenario}
\end{figure}

An \ac{AC} algorithm \cite{konda99} is an off-policy hybrid reinforcement learning that relies on the entities called \textit{actor} and \textit{critic}. The former provides the actions to the transmission system for reward optimization, while the latter estimates the goodness of the choices and sends feedback to the actor. In this work, the \ac{AC} algorithm enables our scheduler to learn the optimal policy for real time traffic allocation given the wireless path conditions and using a reward function that constitutes the \ac{E2E} loss, bandwidth utilization, and the packet loss rate on each path. Figure \ref{fig:scenario} shows the reference scenario we consider in this work, where a \ac{UAV} transmits a multimedia stream towards a \ac{GCS} over multiple wireless paths. The AC-PADME scheduler, relying on a \ac{RTP}/\ac{UDP} protocol stack, determines an optimal policy to select a subset of the paths to be used and the necessary redundancy rate to increase the probability of correctly delivering all packets.  The \ac{AC}-based algorithm allows to find the trade-
problem as an optimization problem and provide an RL		off between the amount of data replicated over the wireless
based solution using an AC algorithm.		channels, for increasing the probability of receiving all packets
• We provide simulation results to evaluate the learning		correctly at the destination, and the amount of used bandwidth.
performance of our algorithm and show that our schedu-		Figure 1 shows an example where packets 1, 2 and 5 are
ler, using the AC algorithm can target very low E2E loss		replicated 2 times, packet 4 once, and packet 3 has no
rate without using excessive bandwidth.		replica. The technique of using redundant packets over a
transmission channel to protect the information is used by
other techniques such as the Forward Error Correction (FEC) \cite{celandroni2011performance}.

The rest of the paper is organized as follows: Section \ref{sec:RelatedWork} gives an overview of the application of \ac{RL} and \ac{AC} to network traffic management and control. The system model is given in Section \ref{sec:methodology}, while in Section \ref{sec:problem_formulation} we formulate the optimization problem and provide an \ac{AC}-based solution. Section \ref{sec:results} discusses the simulation results and Section \ref{sec:conclusion} concludes the paper and points out future research directions.

\section{Related Work}
\label{sec:RelatedWork}
Following the recent advancement in the application of Machine Learning techniques in multimedia transmission, researchers have proposed \ac{RL}-based frameworks for network resource management and control, particularly for traffic scheduling. In \cite{wang2020towards}, a \ac{DQ} \ac{RL}-based scheduler is proposed for bandwidth allocation at a WiFi access point in order to meet the \ac{QoS} requirements of different user applications. Wu et al. \cite{wu2020peekaboo} have proposed Peekaboo, a \ac{RL}-based multipath scheduler implemented in Multipath QUIC to target heterogeneity of both WiFi and cellular channels. In \cite{zhong2019deep}, an \ac{AC}-based framework is proposed for dynamic wireless multi-channel access; the \ac{UE} is trained to select a channel with favourable conditions and avoid collisions. Yang et al.\cite{yang2019actor} have coupled  \ac{AC} framework with a fuzzy normalized radial basis for packet scheduling in Cognitive Internet of Things systems (CIoT) in order to increase channel rate and throughput. In \cite{wei2018joint}, an \ac{AC}-based framework is used to reduce the end-to-end delay in Fog-based IoT systems by optimizing computing, task offloading and resource allocation.

Different from the works presented above, our approach addresses at the same time the two main challenges of multipath transmission; that is, path selection and flow protection through replicas. Our scheduler does not only find the optimal policy for selecting a subset of the available paths but also for determining the required redundancy while keeping a trade-off between flow protection and efficient bandwidth usage. Moreover, the proposed scheduler called AC-PADME is based on the \ac{AC} algorithm that does not require off-line training or apriori knowledge to model the state of the communication system in order to find the optimal policy. The algorithm searches the optimal policy on a parametrized family of functions using a gradient-based approach and leveraging a feedback loop to collect \ac{CSI} on the wireless paths.

The technique of using redundant packets to protect information is also used in \ac{FEC} as described in \cite{bacco2019real, BACCO2022100443, 8647368} and its variants such as BCH \textit{(Bose, Chaudhuri, and Hocquenghem)}, Reed-Solomon and convolution codes \cite{gotta2008experimental, 7870757}. However, different from these \ac{FEC} coding schemes, we employ multipath technique to  distribute the redundant packets over multiple paths, thereby providing resiliency to the unavailability of one or more paths and avoiding coding/decoding overhead as well as retransmission delays.  Recalling that our reference scenario is real-time multimedia delivery, we emphasize how this approach is advantageous in increasing reliability and reducing delay because the expiry time of sent packets is strictly limited.

\section{System Models}
\label{sec:methodology}
\begin{table*}
    \begin{tabular}{|p{24mm}|p{6mm}|p{6mm}|p{6mm}|p{5mm} || p{6mm}|p{12mm}|p{6mm}|p{12mm}|p{6mm}|p{12mm}|p{6mm}|p{12mm}|}
    \hline
    \multirow{3}{*}{\textbf{\centering Scheduling class}} & \multicolumn{3}{c|}{\multirow{3}{*}{\begin{tabular}[c]{@{}c@{}}\textbf{\centering Redundancy ratios}\\{$[r_1, r_2, r_3]$}\end{tabular}}} & \multirow{3}{*}{\centering $\overline{f}$} & \multicolumn{8}{c|}{\textbf{\centering Path Interleaving: The first four rounds of a cycle}}       \\ \cline{6-13} 
                                      & \multicolumn{3}{c|}{}                                                                                              &                    & \multicolumn{2}{c|}{Round 1} & \multicolumn{2}{c|}{Round 2} & \multicolumn{2}{c|}{Round 3} & \multicolumn{2}{c|}{Round 4} \\ \cline{6-13} 
                                      & \multicolumn{3}{c|}{}                                                                                              &                    & Reps         & Paths           & Reps         & Paths            & Reps         & Paths           & Reps         & Paths            \\ \hline
                                      
    \centering G & \centering 0 & \centering 0 & \centering 1 & \centering 3 & \centering 3 & $c_1c_2c_3$ & \centering 3 & $c_1c_2c_3$ & \centering 3 & $c_1c_2c_3$ & \centering 3 & $c_1c_2c_3$ \\ \hline
    
    \centering F & \centering 0 & \centering 0.5 & \centering 0.5 & \centering 2.5 & \centering 3 & $c_1c_2c_3$ & \centering 2 & $c_1c_2$ & \centering 3 & $c_3c_2c_1$ & \centering 2 & $c_3c_1$ \\ \hline
    
    \centering E & \centering 0 & \centering 0.75 & \centering 0.25 & \centering 2.25 & \centering 3 & $c_1c_2c_3$ & \centering 2 & $c_1c_2$ & \centering 2& $c_3c_1$ & \centering 3 & $c_2c_3c_1$ \\ \hline

    \centering D & \centering 0 & \centering 1 & \centering 0 & \centering 2 & \centering 2 & $c_1c_2$ & \centering 2 & $c_3c_1$ & \centering 2 & $c_2c_3$ & \centering 2 & $c_1c_2$ \\ \hline

    \centering C & \centering 0.25 & \centering 0.75 & \centering 0 & \centering 1.75 & \centering 2 & $c_1c_2$ & \centering 2 & $c_3c_1$ & \centering 1 & $c_2$ & \centering 2 & $c_3c_1$ \\ \hline

    \centering B & \centering 0.75 & \centering 0.25 & \centering 0 & \centering 1.25 & \centering 2 & $c_1c_2$ & \centering 1 & $c_3$ & \centering 1 & $c_1$ & \centering 2 & $c_2c_3$ \\ \hline

    \centering A & \centering 1 & \centering 0 & \centering 0 & \centering 1 & \centering 1 & $c_1$ & \centering 1 & $c_2$ & \centering 1 & $c_3$ & \centering 1 & $c_1$ \\ \hline

    \end{tabular}
    \caption{AC-PADME mechanisms: (left) the scheduling classes with redundancy ratios $r_i$;\\(right) the interleaving policy, showing the channels rotation in the first 4 rounds.}
    \label{table:1}
\end{table*}
This paper considers a transmission system that sends $K$ information packets over a set of $m$ wireless paths by using a multipath strategy. We assume that the $K$ packets are sent through a \ac{RB}-based mechanism in  time $T$ called cycle. Within a cycle, the packets are sent using $K$ transmission rounds. The packet transmitted during round $i$ of cycle $j$ can be replicated over a subset of the $m$ wireless paths. A path is defined by a 5 tuple of IP/Port addresses and path identifier. 

We assume that the number of replicas used for each transmission round can be chosen from among predetermined values. These values are evaluated in such a way that the sequence of scheduled packets is feasible for the \ac{RB}-based mechanism. Hence the sequence of replicas used in a cycle defines a scheduling class.  The redundancy ratios and the scheduling class can be related by defining the vector of the redundancy ratios $\textit{\textbf{r}}=[r_1,r_2,\cdots r_m]\;|\;r_i\leq1 \; i=1,\cdots, m;\; \sum_{i=1}^m r_i=1$ and by the following equation $\overline{f}=\sum_{i=1}^m i\cdot r_i$, where $r_i$ is the percentage of packets replicated over $i$ paths  within a cycle and $\overline{f}$ is the average redundancy factor. 

Without loss of generality, to make the concepts introduced above clearer, let us apply them to the scenario analysed in this paper, which relies on a number of wireless paths $m=3$. In this scenario, the left side of table \ref{table:1} shows the predetermined seven scheduling classes from $A$ to $G$, the redundancy ratios vector for each scheduling class $\textit{\textbf{r}}~=~[r_1,r_2,r_3]$, and the average redundancy factor $\overline{f}$. For the scheduling classes defined above, we have a total number of transmitted packets per cycle $N=\overline{f}\:K$ and the redundancy  $R=N-K$. The adopted RB-based mechanism, to work correctly, must be aware of both the redundancy ratios and the number of transmitted packets per round in a cycle.

We need to evaluate the effects of the chosen scheduling class on the communication system through a reward function that consists of the \ac{E2E} packet loss and the bandwidth used to transmit the packets per round. Using the \ac{CSI} from the receiver  feedback as provided by the \ac{RTP}/\ac{RTCP} protocol stack, the packet loss rate per path can be computed as the ratio of the packets lost on the path and the total packets transmitted over the path in an interval time $t'$, as shown in the following equation: 
\begin{equation} \label{eq1}
 PLR{i}(t') = \frac{\sum_{\tau=1}^{t'}PL_{i}(\tau)}{\sum_{\tau=1}^{t'}PT_{i}(\tau)} \; i=1,\cdots,m 
\end{equation}
\noindent where $PT_{i}(\tau)$ and $PL_{i}(\tau)$ are the packets sent and the packets lost on path $i$, respectively,  at time $\tau$. In our scenario $t'$ is the system RTCP report interval.

In the communication between the source (\ac{UAV}) and the destination (\ac{GCS}), the \ac{E2E} packet loss counts the packets that are lost over all the $m$ channels in the time interval $t'$. We can define the \ac{E2E} packet loss as in the following equation:
\begin{equation} \label{eq2}
 E2EPLR(t') = \frac{ \sum^{t'}_{\tau=1} \sum^m_{i=1} PL_i(\tau)}{ \sum^{t'}_{\tau=1} \sum^m_{i=1} PT_i(\tau)}
 \end{equation}
\noindent where the numerator and the denominator count the lost packets and the transmitted packets, respectively, over the $m$ paths  until $t'$ , which is again the RTCP report interval.

Finally, the used bandwidth to transmit the $K$ information packets using  a given scheduling class can be defined as follows:
\begin{equation}\label{eq3}
{Bw} = \overline{f} K
\end{equation}

In the next section we formulate the optimization problem addressed through the \ac{AC} approach involving the system aspects discussed in this section.
\section{Problem Formulation}
\label{sec:problem_formulation}
The optimization problem addressed in this work can be stated as the 
minimization of the \ac{E2E} packet loss using the minimum bandwidth to transmit the information packets using feasible redundancy. 
The state of the communication system that can be observed when performing the minimization is represented by the packet loss rate of the wireless paths.

The optimization problem stated above can be formulated as a decision-making problem that can be modelled as a Markov Decision Process (MDP) \cite{konda99}.
An MDP is defined by the tuple $\{S, A, P(s'|s,a), r(s',s,a)\}$, where $S$ is the states space of the system, $A$ denotes the actions space that can be adopted to perform the optimization, $P(s'|s,a)$ is the transition probability over the state space conditioned by the actions, and $r(s',s,a)$ is the immediate reward due to the action $a$ that triggers the state transition from $s$ to $s'$. In the following, we define the MDP for our problem.
\begin{enumerate}
\item
\noindent$States$: We define the state as the vector $\textit{\textbf{s}} =[ \ell_{1},\ldots,\ell_{m}]$ of the packet loss rate of the wireless paths.
\item
\noindent$Actions$: For the scheduler to minimize the \ac{E2E} using the minimum bandwidth, it has to determine the redundancy factor $\overline{f}$ and the vector of weights $\omega_{i}$ of the \ac{WRR} mechanism. The redundancy factor is computed from the redundancy ratios of the selected scheduling class as discussed in the previous section, while $\omega_{i}$ is the amount of packets to be transmitted over path $i, \; i=1,\ldots, m$. Hence, in our case the action  can be defined as the vector $a = [\overline{f}, \omega_{1},\ldots,\omega_{m}]$
\item
\noindent$Reward:$ In this work, the immediate reward $r(s',s,a)$ provides a penalty $-1$ in the following cases: when the \ac{E2E} loss rate is above a given threshold $\varphi_{th}$; even if the \ac{E2E} loss rate is below the threshold, the penalty is provided for increasing the used bandwidth or if the highest number of packets is not sent on the channel with the lowest packet loss rate:  $\omega_{u}>\ldots>\omega_{m}>\ldots>\omega_{1}>\ldots>\omega_{v}\nRightarrow PLR_{u}<\ldots<PLR_{m}<\ldots<PLR_{1}<\ldots<PLR_{v}$. In the other cases the immediate reward function provides the incentive $1$.
\end{enumerate}

We do not need to define a model for the probability function $P(s'|s,a)$, because as stated previously, we use an \ac{AC} algorithm which can approximate the probability function using the acquired observations of the states without using any apriori model. Moreover, the function policy used to select the actions can be approximated iteratively using a parametrized family of functions. We now clarify how an AC algorithm works by introducing the basic concepts presented by the authors in \cite{konda99,grodman2012}.

\begin{figure}[hbt!]
\centering       \includegraphics[width=.8\columnwidth]{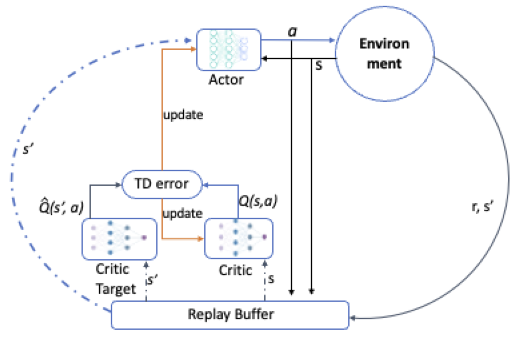}
   \caption{Architecture of the Actor-Critic Learning Algorithm}
    \label{fig:2}
\end{figure}

Figure \ref{fig:2} shows the architecture of the \ac{AC} algorithm. The main entities involved in the algorithm are the critic and the actor, which interact jointly with the system (environment) for achieving the optimization goal. The actor aims to estimate the policy function $\pi(a|s)$ used to select actions based on the state of the environment. This function belongs to a parameterized family of functions; the estimation of the parameters to fit the best one is performed with different techniques such as Neural Networks (NN). The equation used for parameter fitting involves maximizing the expected value of the total reward discounted by the parameter $\gamma$, as shown in the following equation:
\begin{equation} \label{eq4}
        \pi^*(a|s)=\arg \max E\Big[{\sum_{t=0}^{\infty}{\gamma^t(r(s',s, a)\Big]}}
    \end{equation}

The critic evaluates the goodness of the estimated policy function and then sends the feedback to the actor to refine the parameter fitting. The critic performs this operation estimating the value of the state-action function $Q^\pi(s', a)$, solving equation \ref{eq5} that also, in this case, can be solved through a NN.
\begin{equation} \label{eq5}
        Q^\pi(s',a)= E\Big[r(s',s, a) + \gamma(Q^\pi(s,a)\Big]
      \end{equation}
In  figure \ref{fig:2},  the architecture is based on a double critic network. Together with the critic network, there is also a target critic network which is used to estimate the time difference error in order to overcome the instability of the results obtained using the single critic network, as explained in \cite{konda99,grodman2012}. In this work, we apply the soft-update of the target critic network using the critic parameters according to the following equation:
\begin{equation} \label{eq8}
     \theta_{targ.\: crit.}^{new}= \alpha\theta_{targ.\: crit.}^{old} + \theta_{crit.}(1-\alpha) \end{equation}
 where $\theta$ is the TD error for the critic as analysed in \cite{konda99,grodman2012}, and $\alpha$ is a weighting factor that takes into account the estimate from the critic network versus the target critic.

The NNs of both the actor and the critic are designed as Deep NN (DNN) using TensorFlow-2 and Keras libraries, and ADAM algorithm as the underlying optimizer. 
 
 After the transmission of the packets using the selected action $a$, the algorithm evaluates the reward $r(s',s,a)$ and the triggered transition state from $s$ to $s'$, and observes the path loss rates $PLR_i\;i=1,\ldots,m$. The observed data $\{s,s',a,r(s',s,a)\}$ are stored in the replay buffer of size $B$ in order to solve the sample deficiency problem. Once the replay buffer is full, some samples are randomly drawn to update the parametrized models of both the policy and the value-state functions. The updating is based on the Temporal Difference (TD) approach that provides the following loss functions:
  \begin{gather} 
  \label{eq6}        
      \Delta^C =\frac{1}{B} \Big(r(s',s,a)+ \gamma\hat{Q}^\pi(s',a)-Q^\pi(s,a)\Big)^2\\
       \label{eq7}
       \Delta^A = -\delta \ln\pi(a|s) 
  \end{gather}
These functions are used to update the value-state function (critic) and policy function (actor), respectively; where $\delta$ is the TD error for the actor as analysed in \cite{konda99,grodman2012}.

\section{Simulation Results and Analysis}
\label{sec:results}
Simulation results are presented in this section to evaluate the performance of the proposed scheduler in terms of \ac{E2E} packet loss and bandwidth utilization. 
The simulator is built leveraging Gstreamer and its modular plugins designed to handle multimedia streaming. The sender-side of our scheduler runs on a Jetson Nano board transmitting video to a desktop machine via three \acp{NIC}, having as reference the scenario in \cite{BACCO2022100443}.

\begin{figure}[hbt!]
\centering
    \includegraphics[width=0.8\columnwidth]{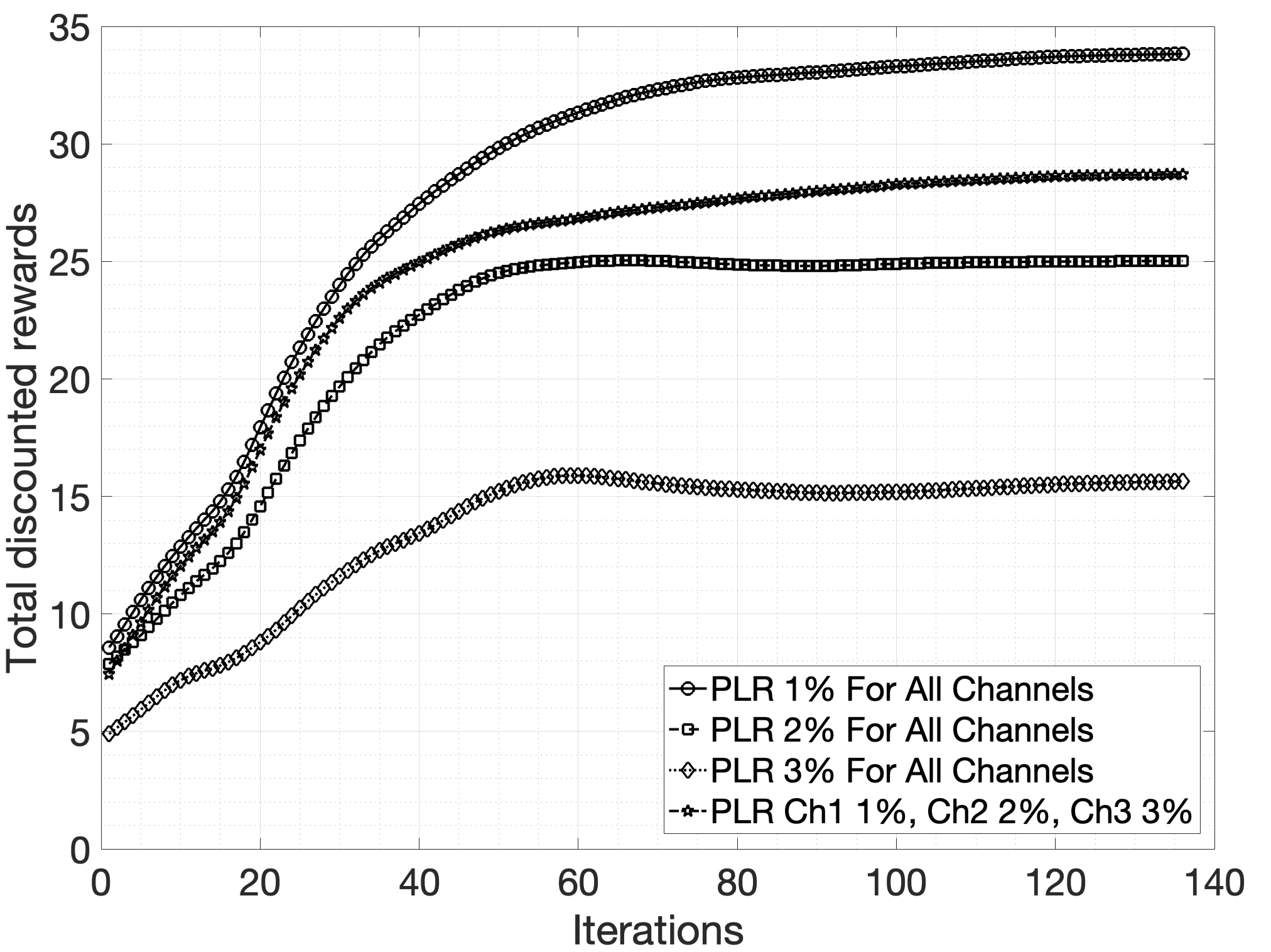}
    \caption{AC-PADME learning performance at different loss rates}
    \label{fig:3}
\end{figure}

Fig. \ref{fig:3} shows the curves of the total discounted reward with different link loss rates. The learning occurs at an interval equal to the RTCP report interval, which in our case is 3s. The figure shows that within $60$  iterations the algorithm reaches a stable state, except in the case with PLR $1\%$ because in this case, the number of loss events useful for estimating models is lower than in other use cases. Note that, in the scenario with different packet loss values per path and a $3\%$ packet loss for all the paths, convergence is achieved before $60$ iterations due to the higher number of loss events available for model estimation. 

\begin{figure}[hbt!]
\centering
    \includegraphics[width=.8\columnwidth]{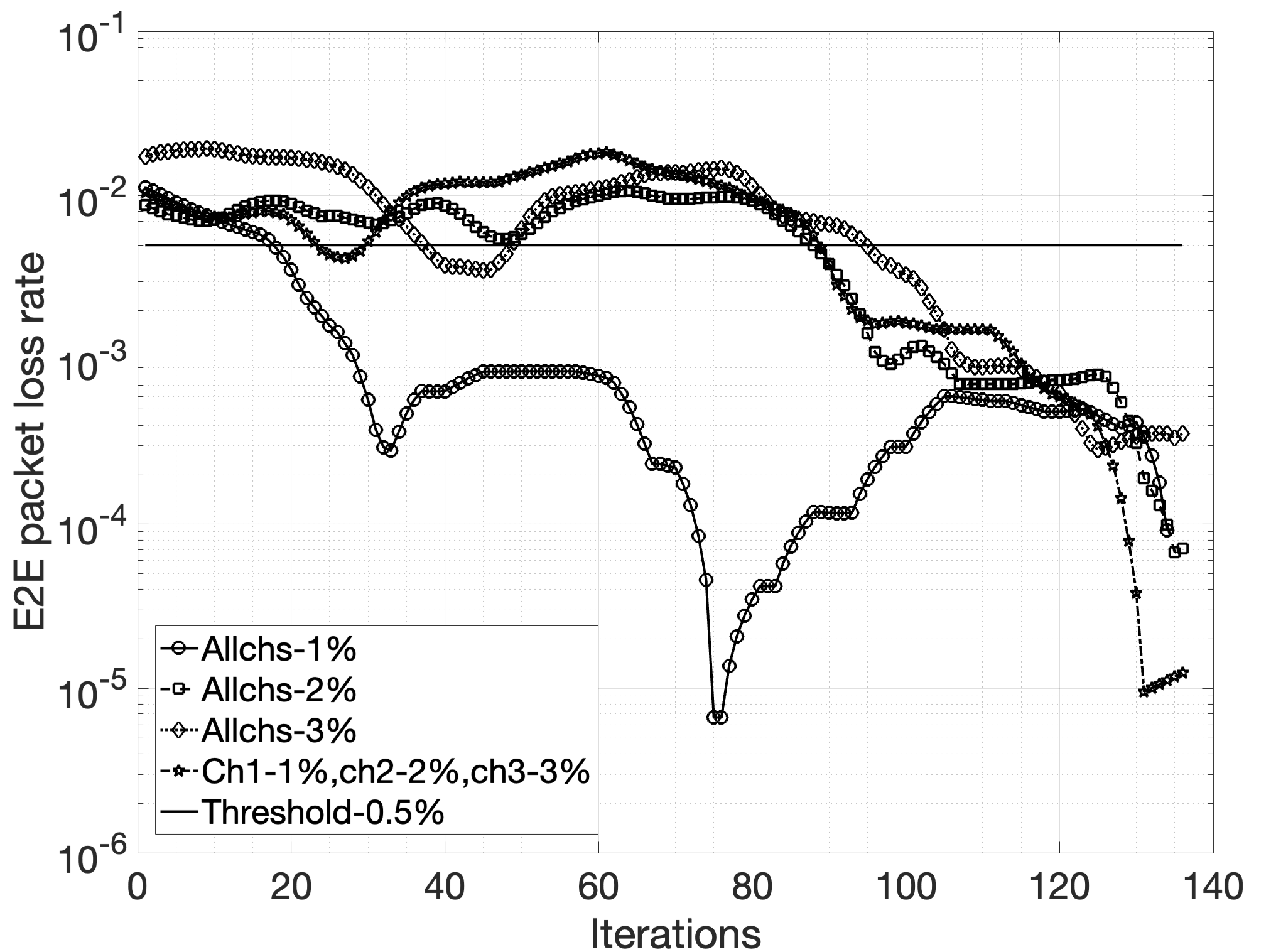}
    \caption{\ac{E2E}-\ac{PLR} with different  path loss rates. The target is $PLR_{th}=0.5\%$}
    \label{fig:4}
\end{figure}

In Fig.\ref{fig:4}, we show the effectiveness of our scheduler in maintaining below a threshold the E2E packet loss. The threshold is set at $0.5\%$ to ensure an optimal QoE in our use case. The figure shows that in the scenario with packet loss $1\%$, the algorithm can guarantee an E2E packet loss below the threshold from iteration $20$ onward, even though, the curve presents a greater variability because the number of available learning events is less than in the other use cases due to the low packet loss rate.   

\begin{figure}[hbt!]
\centering
    \includegraphics[width=.8\columnwidth]{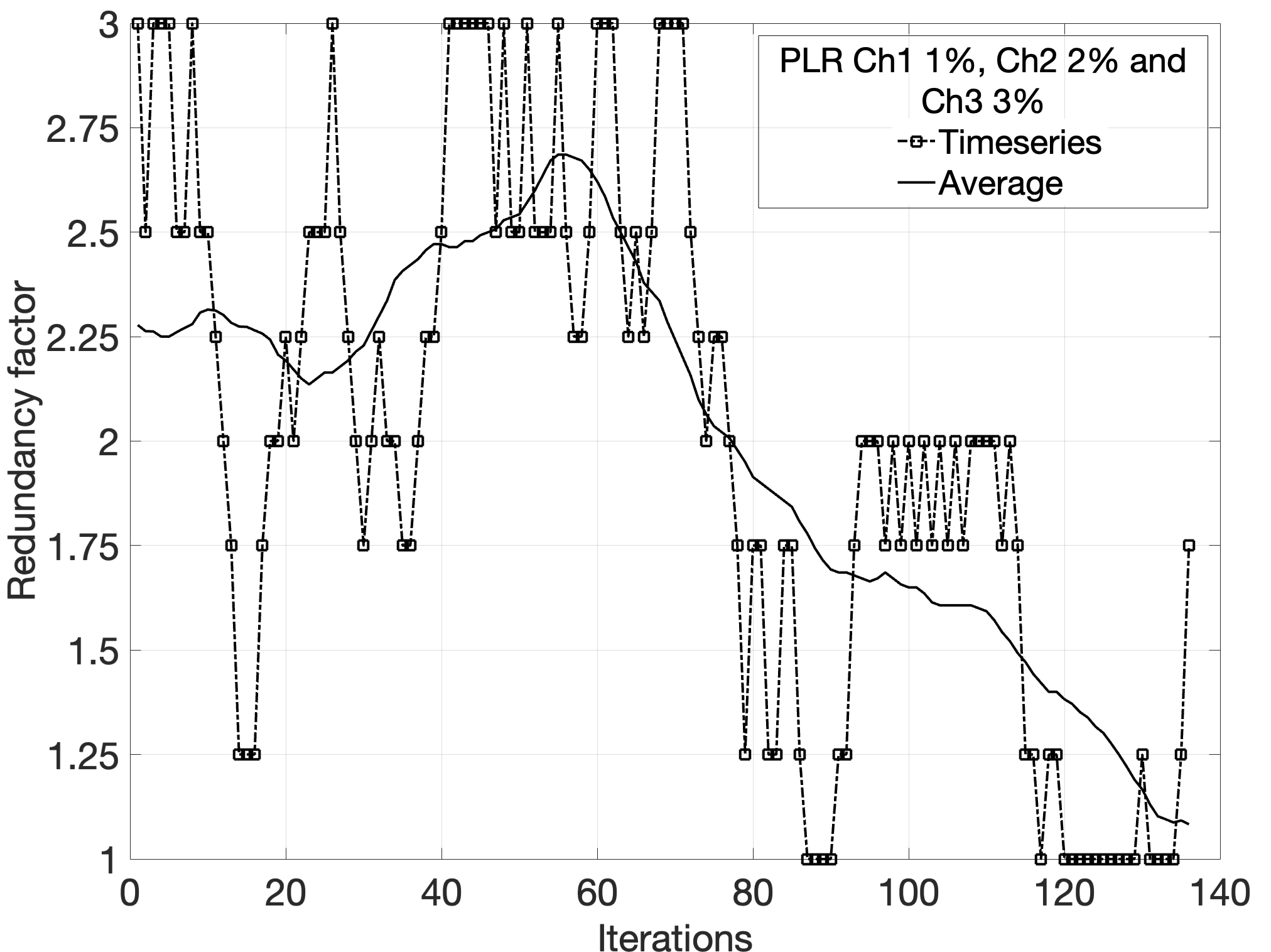}
    \caption{AC-PADME performance in terms of bandwidth utilization at different path loss rates}
    \label{fig:5}
\end{figure}

Fig. \ref{fig:5}  shows the capacity of the scheduler to minimize the bandwidth utilization, while the \ac{E2E} loss is also minimized. The figure shows the different redundancy factors selected by the algorithm during the learning process for the scenario with the paths having heterogeneous packet loss rates . The average redundancy factor curve shows that the bandwidth used reduces rapidly as the algorithm continues to learn.
\section{Conclusions}
\label{sec:conclusion}
In this work, we have proposed an Actor-Critic based scheduler called AC-PADME for multimedia delivery in a multipath environment. The results show that the AC-PADME can progressively learn the scheduling policy by observing the path loss rates and estimating in real-time the required redundancy and the optimal path weight allocation, targeting very low E2E loss, thus improving link availability and the overall \ac{QoS}. These operations are performed through learning procedures that do not need apriori models or training procedures. Our future research will include the analysis of kernel basis functions for the approximation of states space analysed in this work and the possibility to introduce as feedback, the \ac{HIL}, for the tuning of the policy and for improving the \ac{QoE}.

\section*{ Acknowlegments}
This research is supported by TEACHING project funded by the EU H2020 research programme GA n. 871385

\bibliographystyle{IEEEtran}
\balance
\bibliography{bibliography}

\begin{thebibliography}{10}
\providecommand{\url}[1]{#1}
\csname url@samestyle\endcsname
\providecommand{\newblock}{\relax}
\providecommand{\bibinfo}[2]{#2}
\providecommand{\BIBentrySTDinterwordspacing}{\spaceskip=0pt\relax}
\providecommand{\BIBentryALTinterwordstretchfactor}{4}
\providecommand{\BIBentryALTinterwordspacing}{\spaceskip=\fontdimen2\font plus
\BIBentryALTinterwordstretchfactor\fontdimen3\font minus
  \fontdimen4\font\relax}
\providecommand{\BIBforeignlanguage}[2]{{%
\expandafter\ifx\csname l@#1\endcsname\relax
\typeout{** WARNING: IEEEtran.bst: No hyphenation pattern has been}%
\typeout{** loaded for the language `#1'. Using the pattern for}%
\typeout{** the default language instead.}%
\else
\language=\csname l@#1\endcsname
\fi
#2}}
\providecommand{\BIBdecl}{\relax}
\BIBdecl

\bibitem{bacco2017uavs}
M.~Bacco, S.~Chessa, M.~D. Benedetto, D.~Fabbri, M.~Girolami, A.~Gotta,
  D.~Moroni, M.~A. Pascali, and V.~Pellegrini, ``Uavs and uav swarms for
  civilian applications: communications and image processing in the sciadro
  project,'' in \emph{International Conference on Wireless and Satellite
  Systems}.\hskip 1em plus 0.5em minus 0.4em\relax Springer, 2017, pp.
  115--124.

\bibitem{bacco2017survey}
M.~Bacco, P.~Cassar{\'a}, M.~Colucci, A.~Gotta, M.~Marchese, and F.~Patrone,
  ``A survey on network architectures and applications for nanosat and uav
  swarms,'' in \emph{International Conference on Wireless and Satellite
  Systems}.\hskip 1em plus 0.5em minus 0.4em\relax Springer, Cham, 2017, pp.
  75--85.

\bibitem{afzal2019holistic}
S.~Afzal, V.~Testoni, C.~E. Rothenberg, P.~Kolan, and I.~Bouazizi, ``{A
  Holistic Survey of Wireless Multipath Video Streaming},'' \emph{arXiv
  preprint arXiv:1906.06184}, 2019.

\bibitem{konda99}
V.~R. Konda and J.~N. Tsitsiklis, ``{Actor-Critic Algorithms },'' in \emph{Int.
  Conf. NIPS}.\hskip 1em plus 0.5em minus 0.4em\relax MIT Press, 1999, pp.
  1008--1014.

\bibitem{celandroni2011performance}
N.~Celandroni and A.~Gotta, ``Performance analysis of systematic upper layer
  fec codes and interleaving in land mobile satellite channels,'' \emph{IEEE
  Transactions on Vehicular Technology}, vol.~60, no.~4, pp. 1887--1894, 2011.

\bibitem{wang2020towards}
Q.~Wang, T.~Nguyen, and B.~Bose, ``Towards adaptive packet scheduler with
  deep-q reinforcement learning,'' in \emph{2020 International Conference on
  Computing, Networking and Communications (ICNC)}, 2020, pp. 118--123.

\bibitem{wu2020peekaboo}
H.~Wu, {\"O}.~Alay, A.~Brunstrom, S.~Ferlin, and G.~Caso, ``Peekaboo:
  Learning-based multipath scheduling for dynamic heterogeneous environments,''
  \emph{IEEE Journal on Selected Areas in Communications}, vol.~38, no.~10, pp.
  2295--2310, 2020.

\bibitem{zhong2019deep}
C.~Zhong, Z.~Lu, M.~C. Gursoy, and S.~Velipasalar, ``A deep actor-critic
  reinforcement learning framework for dynamic multichannel access,''
  \emph{IEEE Transactions on Cognitive Communications and Networking}, vol.~5,
  no.~4, pp. 1125--1139, 2019.

\bibitem{yang2019actor}
H.~Yang and X.~Xie, ``An actor-critic deep reinforcement learning approach for
  transmission scheduling in cognitive internet of things systems,'' \emph{IEEE
  Systems Journal}, vol.~14, no.~1, pp. 51--60, 2019.

\bibitem{wei2018joint}
Y.~Wei, F.~R. Yu, M.~Song, and Z.~Han, ``Joint optimization of caching,
  computing, and radio resources for fog-enabled iot using natural
  actor--critic deep reinforcement learning,'' \emph{IEEE Internet of Things
  Journal}, vol.~6, no.~2, pp. 2061--2073, 2018.

\bibitem{bacco2019real}
M.~Bacco, P.~Cassar\'a, A.~Gotta, and V.~Pellegrini, ``{Real-Time Multipath
  Multimedia Traffic in Cellular Networks for Command and Control
  Applications},'' in \emph{2019 IEEE 90th Vehicular Technology Conference
  (VTC2019-Fall)}, 2019, pp. 1--5.

\bibitem{BACCO2022100443}
\BIBentryALTinterwordspacing
M.~Bacco, P.~Cassarà, and A.~Gotta, ``Air-to-ground real-time multimedia
  delivery: A multipath testbed,'' \emph{Vehicular Communications}, vol.~33, p.
  100443, 2022. [Online]. Available:
  \url{https://www.sciencedirect.com/science/article/pii/S2214209621001121}
\BIBentrySTDinterwordspacing

\bibitem{8647368}
F.~Gabriel, J.~Acevedo, and F.~H.~P. Fitzek, ``Network coding on wireless
  multipath for tactile internet with latency and resilience requirements,'' in
  \emph{2018 IEEE Global Communications Conference (GLOBECOM)}, 2018, pp. 1--6.

\bibitem{gotta2008experimental}
A.~Gotta and P.~Barsocchi, ``Experimental video broadcasting in dvb-rcs/s2 with
  land mobile satellite channel: a reliability issue,'' in \emph{2008 IEEE
  International Workshop on Satellite and Space Communications}.\hskip 1em plus
  0.5em minus 0.4em\relax IEEE, 2008, pp. 234--238.

\bibitem{7870757}
A.~Badr, A.~Khisti, W.-T. Tan, and J.~Apostolopoulos, ``Perfecting protection
  for interactive multimedia: A survey of forward error correction for
  low-delay interactive applications,'' \emph{IEEE Signal Processing Magazine},
  vol.~34, no.~2, pp. 95--113, 2017.

\bibitem{grodman2012}
I.~Grondman, L.~Busoniu, G.~A.~D. Lopes, and R.~Babuska, ``A survey of
  actor-critic reinforcement learning: Standard and natural policy gradients,''
  \emph{IEEE Transactions on Systems, Man, and Cybernetics, Part C
  (Applications and Reviews)}, vol.~42, no.~6, pp. 1291--1307, 2012.

\end{thebibliography}

\end{document}